\documentclass[preprint]{aastex}

\begin{document}

\title{Star formation and Asymmetry in the spiral arms of M51
--  Variable star formation caused by more than one spiral density wave
}

\author{
Alaina L. Henry, 
A.~C. Quillen \& Robert Gutermuth
\affil{Department of Physics and Astronomy, 
University of Rochester, Rochester, NY 14627; 
{\it alaina@astro.pas.rochester.edu, aquillen@pas.rochester.edu, rguter@astro.pas.rochester.edu }
 }
}

\begin{abstract}

In the inner 3kpc of M51 
we find that logarithmic spirals provide good fits to the peak intensities
in molecular gas observed by BIMA in the CO (J=1-0) emission line
along the spiral arms. However, we measure significant asymmetries between
the location and density of the arms observed on one side of the galaxy
compared to those on the opposite side.  Between a radius of 1 and 2.2 kpc
the gas distribution traced in CO is lopsided with densities twice
as large in one arm compared to the opposite one. In the same
region, the spiral arms are offset by 20 degrees from the position of the arm
on the opposite side of the galaxy after a rotation of 180 degrees.
We use the ratio of CO emission to that in the Pa$\alpha$ 
hydrogen recombination line to estimate
the efficiency of star formation along each arm.  Except for
a narrow region at about 2kpc where star formation is surprisingly
inefficient, the gas depletion time is approximately 200 million years despite
large variations in the molecular gas density in the arms.

We account for the deviations from two armed bi-symmetric structure
with a simple model that includes an additional 
three-armed spiral density wave.   This model accounts for the angular offset between the arm on one side compared to its opposite,
the lopsided distribution in molecular gas, and inter-arm star formation
to the north east of the galaxy nucleus.
Since the star formation efficiency is unaffected by the variations
in gas density, and the variations in gas density can be accounted
for by the presence of an additional three-armed density wave,
we suggest that the star formation rate is variable,
and is highest where and when 
the maxima of the two spiral density waves coincide or constructively add.
The inner region of M51 provides good evidence for the presence of
more than one spiral density wave and a resulting variable rate of 
star formation.

\end{abstract}

\section{Introduction}

The evolution of galaxies is influenced by how quickly gas is consumed 
by the formation of stars.  Many empirical star formation recipes
assume that the star formation rate depends on the gas surface density 
(e.g., \citealt{Kennicutt98}).  However,  observations of 
spiral galaxies show that the molecular gas is concentrated 
along spiral arms, implying that star formation is triggered by the 
spiral structure itself.  The strongest evidence for the important
role of spiral structure is based on a gravitational stability
threshold, below which star formation ceases \citep{Kennicutt89}. 
Nevertheless, above this threshold, azimuthally averaged data have
shown that these empirical star formation laws are obeyed 
(\citealt{Wong02}, \citealt{Martin01}).  
In this paper we make use of recent high
resolution mapping to estimate the rate and efficiency of star formation 
along individual spiral arms.  We aim to investigate the influence
of the spiral structure itself 
on the process of star formation.

M51 is a magnificent example of a nearby grand design spiral galaxy.  
Because of its large angular size and strong two-armed spiral structure, it 
has been the focus of studies of spiral density wave driven star formation 
(e.g., \citealt{Tilanus91}, \citealt{rand92}).
While the stellar density contrast in the spiral arms
in M51 is high compared to many spiral galaxies \citep{rix93}, it is
still far below the density contrasts seen in molecular or ionized gas.
The distribution of molecular gas and associated star formation
are particularly sensitive to the underlying spiral density waves 
propagating through the galaxy.   \citet{rix93} in their Fourier analysis
of the K-band light distribution measured significant deviations
from bi-symmetric ($m=2$) symmetry in the galaxy.  Because
shocks occur in molecular gas, 
even weak perturbations in the stellar density distribution can cause large
density variations in the molecular gas distribution.
By measuring the 
shape and location of the peak densities seen in molecular gas, and
comparing the shapes of the spiral arms with the star formation
rates measured from the Pa$\alpha$ emission, we 
probe for differences in 
star formation rate and efficiency that might be caused
by the spiral density waves themselves.

In this paper we take advantage
of recent high angular resolution observations of M51 
to measure the star formation efficiency along individual spiral arms.  The BIMA 
Survey of Nearby Galaxies (BIMA-SONG, \citealt{helfer}) 
is the first systematic interferometric survey of the disks of nearby galaxies in 
the CO(J=1-0) emission line.  Prior to BIMA-SONG, the overwhelming majority 
of galaxies observed in CO had been
observed with single dish telescopes, at much lower angular resolution.  
With a typical beam full width half max of $\sim 6''$, BIMA-SONG represents a major improvement in our ability to resolve the molecular gas 
distribution in a significant sample of nearby galaxies.      

To measure the gas depletion time, which is inversely related 
to the star formation efficiency, we 
combine a measurement of the gas density with an estimate of the star formation
rate.  To measure the star formation rate, 
we use archival NICMOS data in the Pa$\alpha$ recombination line, at 1.87$\mu$m.
The Pa$\alpha$ recombination line is better suited toward
probing the possibly obscured central regions, 
since the near infrared is less affected by dust extinction. The 
high angular resolution of the NICMOS images on board
HST facilitates continuum subtraction in crowded regions.

In Section 2, we describe the archival observations used for this study.  
To measure quantities along
spiral arms, we must first determine their locations.  
In section 3.1 we describe our procedure for doing this and discuss
our measurements that describe the shape of the arms.  
In section 3.2 we discuss 
the relative differences in gas density, star formation rate
and gas depletion time measured in different spiral arm
regions.  In section 3.3 we discuss the effect of a possible additional
spiral density wave on the gas density and on the symmetry of the spiral arms
previously identified.  A summary and discussion follows.

For M51 we adopt a distance of 7.7 Mpc \citep{Tully}, so $1''$ 
on the sky corresponds to a distance of 37 pc.  

\section{Observations}

NICMOS narrow band (F187N and F190N) images were obtained as part of the NICMOS
GTO program \citep{scoville} with a mosaic of nine fields covering
an area of about $2'\times 2'$.    
See \cite{scoville} for more information about these observations.
The F187N and F190N filters, with
effective wavelengths of 1.87 and 1.90$\mu$m contain Pa$\alpha$
line and continuum images, respectively.  
We scaled and subtracted the F190N continuum images
from the F187N images and combined them to form one mosaic. 
The pixel scale of the NICMOS images is $0.204''$, so one pixel is 7.6 pc long in M51.  
The NICMOS images were calibrated using zero points listed in
the HST Data Handbook.  Specifically, we used $8.723\times 10^{-14}$
erg~cm$^{-2}$~s$^{-1}$/(DN~s$^{-1}$) for the continuum subtracted F187N narrow band images.

The BIMA-SONG database contains a collection of spatial velocity data cubes.  
For this study, we used an image made of
the intensity integrated over all velocity channels. 
For M51, large scale fluxes were recovered
in the interferometric data from single dish observations \citep{helfer}.   

To compare the CO image to the Pa$\alpha$ image we must ensure that the same 
areas of the sky can be measured on the same scale.  
Therefore,  we resampled the CO image to be at the same pixel scale
as the Pa$\alpha$ mosaic. 
Both images are inclination corrected 
 and rotated so that the minor axis is at a P.A. of 170 degrees.  
 The resulting images are shown in
Figures 1 and 2.  

\section{Results}

\subsection{Logarithmic Spirals}

To determine the location of the spiral arms, we 
model the spiral density waves in the disk with the following function,
that describes a logarithmic spiral  
\begin{equation}
\Sigma \propto 
{\rm Re} \left\{  e^{i({\alpha_m \ln r - m (\phi  - \phi_{0}) })} \right\}  
\end{equation}  
where $\alpha$ describes how tightly wound the arm is, $\phi_{0}$ is an angular offset,
and $m$ is an integer that corresponds to the number of spiral arms. 
In the case of M51, the spiral structure is predominantly two armed ($m=2$).
One maximum is then described by
\begin{equation}
\label{logparam}
\alpha_m \ln r = m (\phi + \phi_{0}).
\end{equation}
For each region we measured parameters $\alpha_2$ and $\phi_{0}$ by choosing 
several positions of peak intensity along
the spiral arms in the CO intensity map. All positions
were measured from the center of the galaxy,
which was determined from the centroid of an H band (F160W) image of M51 that was observed simultaneously with the Pa$\alpha$ data
\citep{scoville}.  
Since the nuclear region of M51 contains little molecular gas (see Figure 1), 
the H band centroid accurately measures the location of the stellar density
peak.  We defined the angle $\phi = 0$ to be toward the right in Figures 1 and 2.     
From these measurements, sets of coordinates, ($r, \phi$), were measured in each arm.  
Linear regression was used to determine $\alpha_2$ and $\phi_{0}$ in each region.  
Different regions in the spiral arms were fit separately,
as a single spiral did not provide an adequate fit to the entire galaxy. 
The fits to the 8 selected spiral sub-regions are illustrated in Figure 1, 
and the corresponding parameters that describe them, are listed in Table~1.  

In Figure 2 we compare our fit to the spiral arms traced in CO to the Pa$\alpha$ emission.    
While the Pa$\alpha$ is coincident with the 
molecular gas in some of the spiral arms, in other regions an offset is observed.  
\citet{Tilanus91} and \citet{rand92} noted 
an offset between CO and H$\alpha$ in the sense that H$\alpha$ emission was 
at somewhat larger radii than the CO emission.   
We find that the Pa$\alpha$ emission is offset from the CO emission at radii 
1.2 to 2.8 kpc in the spiral arms 2b and 2c. 
However, we cannot tell if there is
an offset between the Pa$\alpha$ emission and CO emission in spiral arm 1c because
the edge of the Pa$\alpha$ image is near the position of the CO emission.
We do note hints of 
Pa$\alpha$ emission, to the south of spiral arm 1c, lying on the 
very edge of the NICMOS mosaic. If there is Pa$\alpha$ emission to the south of the 
mosaic, then a significant offset between the CO and Pa$\alpha$ is present in this arm.  

In addition to observing offsets between CO and Pa$\alpha$, 
we can look for deviations from point symmetry by plotting the
reflection of the spiral arms 
determined in Table 1.  In Figure 2 we show the fit to the spiral arms, 
rotated through an angle of 180 degrees.   In the case of perfect $m=2$ 
symmetry, the over-plotted curve would lie on top of the peak
intensities.   We observe that the most inner and outer spiral arm regions 
appear to be symmetric, but for intermediate
radii, the observed spiral arms are offset from the curves.  To verify this, 
in Figure \ref{offsets} we plot $r$ as a function of $\phi$ for the 
a, b, and c spiral 
arms listed in Table 1, with the northern spiral arm rotated 
by an angle of 180 degrees.  We note
that the deviation from point symmetry is present 
over much of the inner part of the galaxy, but is most apparent for the
b spiral arms, where the offset, $\Delta \phi \sim  20$ degrees.

\subsection{Star Formation Rates and Gas Depletion Times}

In this section we present our measurements of the star formation rate (SFR) 
and CO mass, per unit length, along each segment
of the spiral arms.  Figures 4, 5, and 6 show the molecular gas 
density in each spiral arm segment and the star formation rate per unit length along 
the spiral arm segments.  In addition, we calculated the gas depletion time 
from the ratio of the CO mass to the star formation rate.  
To measure these quantities, we used our own IDL procedure to measure the fluxes
in particular regions.  Each region was centered in radius on the position of
a logarithmic spiral curve determined from our study described
in section 3.1 (see Figures 1,2).
The regions were defined by intervals in $\phi$ and $r$, with 
$\Delta \phi \sim  0.05$ radians and $\Delta r \sim   50$ pixels 
or approximately 400 pc.  The same 
regions were used to measure fluxes in both the CO image and the Pa$\alpha$ image, so that
a one to one comparison could be made at each location.
We ensured that $\Delta r$ was large enough to include the 
Pa$\alpha$ emission, where offsets between Pa$\alpha$ and
CO were observed. 
The measured fluxes were divided by arc length, 
so that integrated fluxes were computed per pc along each arm. 
Background sky values were determined in the Pa$\alpha$ image by taking the
median of the regions adjacent to each box, on both sides of the spiral arms.
Only spiral arm regions 1a, 1b, 2a, 2b, and 2c are compared, 
as regions 1c, 1d, and 2d do not lie on the NICMOS Pa$\alpha$ mosaic.  

We next convert to physical units assuming an intrinsic flux ratio of
$F_{H\alpha}/F_{Pa\alpha} = 8.46$ for Case B recombination
at a temperature of $10^4$K and a density $100$ cm$^{-3}$
(line ratios are given in \citealt{osterbrock}).
Then we converted the H$\alpha$ fluxes to an star formation
rate, using a factor from \cite{Kennicutt94},   
\begin{equation}
{SFR(M_{\sun}~ {\rm yr}^{-1})} 
={ {L_{H_{\alpha}}} \over {(1.25 \times 10^{41} {\rm ergs~ s}^{-1})}}.
\end{equation}
The CO data were converted to M$_{\sun}$ pc$^{-2}$
using a CO to
H$_2$ conversion factor of 
$2 \times10^{20}$ H$_2$ cm$^{-2}$ (K km/s)$^{-1}$ \citep{strong}.  

In Figures \ref{gdt1}, \ref{gdt2},
and \ref{gdt3}, we show the molecular gas mass, star formation rate, and gas depletion time, per unit length, 
for arms covered by both images.  
The plots are arranged such that the top plots show the 
molecular gas mass per pc, along the spiral arm segment.  The middle plots
show the star formation rate per pc, as measured from the Pa$\alpha$ mosaic.  
The bottom plots show the ratios of the points in the two plots above them.  
This quantity is known as the gas depletion time, the time 
in which all of the molecular gas would form into stars, if the star formation
rate was to remain constant.   
In Figure \ref{gdt1}, spiral arm 1a was placed next to 
its opposite spiral arm, 2a, and the same is true for arms 1b and 2b, in Figure \ref{gdt2}.
In Figure \ref{gdt3},  spiral 2c is shown alone, because we
were unable to measure Pa$\alpha$ emission for spiral arm 1c.  
Slightly negative star formation rates, the result of imperfect sky subtraction in the 
Pa$\alpha$ image, are assumed to be sites where essentially no stars are being formed.    

Figures \ref{gdt1} and \ref{gdt2} can be used to look for  differences
in the gas density and star formation efficiency 
between a spiral arm and its partner on
the opposite side of the galaxy.
In Figure \ref{gdt1} we note that the gas density is twice as high on NW side (1a) than
the  SE side (2a).  However, the star formation efficiency is about the same 
with a gas depletion time $\sim  2\times 10^8$ years.  
In Figure \ref{gdt2} we note a similarity, in the sense that the 
gas density is again twice as high in the western spiral arm (1b) as in 
the eastern spiral arm (2b).  Both spiral arms 1b and 2b have a 
jump in gas depletion time (a decrease in star formation efficiency), at a radius of
1.5 kpc.  At radii smaller than 1.5 kpc, the gas depletion time 
in spiral arm regions 1b and 2b is similar to that in 1a and 2a.  
At larger radii the gas depletion time is several times larger than within
this radius.  Star formation in the outer parts of spiral arms 1b and 2b is
inefficient compared to that in the inner parts of the same arms.
Past a radius of 2 kpc the gas depletion time drops back to $2\times 10^8$     
years, and remains similar in spiral arm 2c.  
We suspect that the star formation efficiency in arm 1c would be similar to that
in 2c, although we are unable to observe it.    

Except for the outer parts of spiral arms 1b and 2b at a radius
of about 1.5 to 2kpc, the star formation
efficiency or gas depletion time is remarkably similar in all arms,
and approximately $2 \times 10^8$~yr,
despite large variations in gas density and the star formation rate.
In the outer parts of 1b and 2b the gas depletion time is
longer, $\sim 10^9$ years, despite the high molecular
gas densities in these arms.

\subsection{Deviations from $m=2$ symmetry}

In this section we consider a dynamical model that can explain
the degree of deviation from $m=2$ symmetry that we have measured
in the spiral arms of M51.
By expanding galaxy images in Fourier components,
\citet{elmegreen92} noted that many galaxies exhibit
hidden three armed components and consequently suggested that multiple
spiral density waves often propagate simultaneously in galaxy disks.
\citet{rix93} in their Fourier expansion of the near-infrared images
of M51 saw both $m=1$ and $m=3$ components as well as the
dominant 2-armed structure.  \citet{rix93}
found that the galaxy was lopsided at all radii but also
measured a weak peak in the $m=3$ component of the K-band images
at a radius of about $35''$.  This radius
is approximately at the location where we have found the largest
asymmetry in the spiral structure, although from Figure 3 we note that 
the asymmetry is present throughout most of the galaxy. 

We now consider the possibility that more than one spiral density wave pattern 
is present
in the disk of M51.  Below we estimate the likely position of
molecular gas along spiral arms resulting from more than
one perturbation.
We assume that the perturbations in the gravitational potential
resulting from spiral modes can
be described in terms of Fourier components
\begin{equation}
\Phi(r,\phi) = \sum_m A_m
\cos { (m \phi - \alpha_m \ln{r} - \beta_m)}
\end{equation}
where $A_m$ and $\beta_m$ are amplitude and phase shift, and 
$ m\phi - \alpha_m \ln r$ 
describes an $m$ armed logarithmic spiral (see section 3.1).  
Because spiral density waves travel at pattern speeds
above the sound speed in the interstellar medium, the passage
of spiral modes in the disk causes shocks in the gas.
The concentration of molecular clouds along spiral arms
is interpreted to be a result of these galactic scale shocks.
Shocks in the gas are likely to occur near maxima in
the stellar density field, approximately corresponding to minima 
in the gravitational potential.
These are found at angles where $d\Phi/d\phi =0$ or
\begin{equation}
\label{extreme}
\sum_m A_m  m
\sin { ( m \phi - \alpha_m \ln{r} - \beta_m)}
=0
\end{equation}
This equation is also valid at potential maxima where shocks will
not occur.
We assume that one spiral component dominates;  in the case of M51 this would
be the $m=2$  or two-armed mode.
In the absence of other modes, the potential minima (density maxima)
occur near angles
\begin{equation}
\label{phimaxdef}
2 \phi_{2,max} =  2 n \pi +
 \alpha_2 \ln{r} + \beta_2
\end{equation}
where $n$ is an integer.
We expand  Equation \ref{extreme} about the angle $\phi_{2,max}$, for
$ \phi =  \phi_{2,max} + \delta \phi$, where the offset, $\delta\phi$, is 
the difference between the location of density maximum caused by only one $m=2$ wave
and that resulting from two spiral density waves. 
To first order in $\delta \phi$,
\begin{equation}
4 A_2   \delta \phi
+ A_m m \left[ \sin { ( m \phi_{2,max} - \alpha_m \ln{r} - \beta_m)}
 +  \cos { (m \phi_{2,max} - \alpha_m \ln{r} - \beta_m)}  m  \delta \phi \right]
= 0
\end{equation}
where we have assumed only one spiral density wave with $m \neq 2$,
in addition to the dominant two-armed one.
We solve for $d \phi$
\begin{equation}
\delta\phi =
- {
A_m m  \sin { ( m \phi_{2,max} - \alpha_m \ln{r} - \beta_m)}  \over
4 A_2    +
   A_m m^2 \cos { ( m \phi_{2,max} - \alpha_m \ln{r} - \beta_m)}
}.  
\end{equation}
Substituting for $\phi_{2,max}$ (Equation \ref{phimaxdef}) and taking
the limit $A_m < A_2$
\begin{equation}
\label{dphi}
\delta\phi \approx
- { A_m m
\over 4 A_2 }
\cos{(n m \pi)}
\sin \left(
{ ({m \alpha_2\over 2}  -\alpha_m)\ln{r} + {m\beta_2 \over 2}  - \beta_m}\right).
\end{equation}

We use our model to estimate the angular offset that would be measured
between a spiral arm and its opposite one after a rotation of 180 degrees.
For a dominant two-armed mode, each arm has a different integer $n$ 
in Equation (\ref{dphi}); $n=0$ or $1$.
Only when $m$ is odd will the angular offsets be asymmetric,
causing reflection symmetry about the origin to be broken.
The angular offset between one spiral arm and the other that has been
rotated about $\pi$ is then given by subtracting the $n=0$ offset from
that of the opposite one ($n=1$),
\begin{equation}
\Delta \phi \approx
 { A_m m \over 2 A_2 }
\sin  \left(({ m \alpha_2\over 2}  -\alpha_m)\ln{r} + {m \beta_2\over 2}  - \beta_m \right)
\label{angular_offset}
\end{equation}
for odd $m$. 
We expect that the angular difference depends upon the amplitude
of the additional spiral density wave  and on its wave number.
The angular offset should increase as the amplitude of the additional
spiral density wave increases.  
In a region where the amplitudes of the two dominant
spiral density waves are constant, 
we expect that the angular offset would be more open
(less tightly wound) than the spiral density waves themselves.

We now relate the size of the angular offset to the one expected
based on observed density perturbations or observed Fourier components.
In the WKB approximation
$A_m \sim 
{-2 G \Sigma_m r/  |m\alpha_m |}$
where $\Sigma_m$ is magnitude of the $m$th Fourier component of the mass
surface density \citep{B+T}, so that
\begin{equation}
{A_m \over A_n} \sim
{ \Sigma_m n |\alpha_n| \over \Sigma_n m |\alpha_m |}.
\end{equation}
Substituting this into equation(\ref{dphi}) and
using the ratios ${\Sigma_3 \over \Sigma_0}\sim 0.05$,  ${\Sigma_2 \over \Sigma_0}\sim 0.15$, 
and ${\Sigma_1 \over \Sigma_0}\sim 0.15$ at $r=35''$
measured by \citep{rix93} we estimate
\begin{eqnarray}
\label{dphiapprox}
\Delta\phi \approx
\sin  \left(  ({ \alpha_2\over 2} -\alpha_1)\ln{r} + { \beta_2\over2} - \beta_1\right) &{\rm for}~m=1 \nonumber \\
\Delta\phi \approx 0.3
\sin  \left(  ({3 \alpha_2\over 2} -\alpha_3)\ln{r} + {3 \beta_2\over 2} - \beta_3\right) & {\rm for}~m=3  
\label{offset_eqn}
\end{eqnarray}
where the angles are given in radians.
Assuming $\alpha_2 \sim  \alpha_m$, and for a maximum value of the angular offsets
($\beta_2$ with respect to $\beta_m$) the angular shift could be as
large as 60 degrees for $m=1$, and 20 degrees for $m=3$. 

We compare the estimated model angular offsets 
to those we measured in M51.
Recalling from Figure \ref{offsets}, the offset measured 
between the 1b and 2b spiral arms is
$\Delta \phi \sim 20$ degrees.  So the $m=3$ Fourier component observed
by \citet{rix93} is sufficiently large to account for the angular offset
we measured between these two arms.
A three-armed mode, in addition to the dominant two-armed one, consistent
with perturbations seen in the K-band Fourier components, 
could explain the asymmetry
in the locations of the two spiral arms seen in molecular gas.  
Figure \ref{offsets} shows that 
while the asymmetry is present over much of the galaxy, 
the maximum offset occurs where \cite{rix93} observe a peak in the $m=3$ mode.  
The $m=1$ mode, on the other hand, is strong over much of the entire galaxy and
does not single out any particular radial region,  though it could be  
responsible for asymmetry at larger and smaller radii, 
where the $m=3$ amplitude is weak.       
The decrease of the angular offset at small and large radii ($r<1.2$kpc and $r>2.3$kpc) 
can be explained by the decrease in amplitude of the $m=3$ wave.   The nearly
constant value of the angular offset 
between $1$ and $2.2$ kpc suggests that the wave numbers of the $m=2$ and $m=3$
modes are approximately the same size 
(${3 \alpha_2\over2} \sim \alpha_3$, see equation \ref{offset_eqn}).


If there is both $m=2$ and $m=3$ waves, then at least 
one density peak of the $m=3$ wave will lie in between the dominant $m=2$
arms.  We might expect inter-arm molecular gas and associated inter-arm 
star formation in the same region where we see the deviations from bi-symmetry.    
A more careful inspection of Figure 1 shows that there is a patch of interarm
star formation within spiral arm 2b to the north east of the nucleus.
As we have done for the molecular gas distribution (see Section 3.1), 
we crudely fit a spiral shape to the
the Pa$\alpha$ emission in this region.  The resulting fit is shown in Figure \ref{tri},
along with the same arm rotated by 120 and 240 degrees.
From this figure, we can determine where the $m=3$ and $m=2$ waves are likely
to constructively add and where they interfere.
The $m=3$ wave is coincident with the dominant 
spiral arm in the west (1a-1b), but is offset from the spiral arm 
in the southeast (2a-2b).  We find that 
the $m=3$ wave can account for the molecular gas density
enhancement in the western spiral arms, relative to the eastern spiral arms 
(see Figures \ref{gdt1}, \ref{gdt2}, and \ref{gdt3}).  

The addition of an extra wave should also introduce an asymmetry in the velocity
field.  If this asymmetry is sufficiently large then
it could be detected in the CO velocity field.  
To search for this 
we examined the mean velocity MOM1 BIMA map of M51.
We first subtracted the systemic
velocity of the galaxy from the velocity field and then compared this field
to the same field that has been rotated by 180 degrees and multiplied by $-1$.
We find that the line of sight velocity on the southern side of 
the galaxy  2kpc from the nucleus is about 20-30 km/s 
above that on the opposite side.  The velocity asymmetry seems to be largest
in the same region where we see the other asymmetries.   
The magnitude of the velocity asymmetry should be something that
our simple model could predict.  
However, it depends on the pattern
speed of the two waves and the details of the
gas dynamics so a simulation is probably required to make
a prediction sufficiently accurate to compare to the complex
streaming motions seen in the CO channel maps.


\section{Summary and Discussion}

In this paper we have fit the spiral arms of M51 as traced in CO 
within 3kpc of the nucleus with logarithmic spirals.  
We have detected an asymmetry
between the locations of the arms on one side of the galaxy compared
to those on opposite side.  The asymmetry corresponds 
to a maximum angular offset of about $20$ degrees  between one arm 
and the opposite one rotated by 180 degrees, 
and is most prominent between 1 and 2.2 kpc from the nucleus.
The molecular gas density in this region is also asymmetric with one spiral arm 
about twice as dense as that on the opposite side.  In the same radial
region, there is a patch of star formation (seen in the Pa$\alpha$ images
located between
the dominant spiral arms to the north east of the galaxy nucleus.

To explain the angular offset seen between one arm and the opposite one rotated
by $\pi$, we consider a model that contains two spiral density waves; a weak
$3$-armed wave in addition to the dominant two-armed one.
When the additional wave
is odd-armed, we predict that the position of the two dominant arms deviate
from point symmetry in the manner we have observed in M51.  We find that an additional $m=3$ mode,
consistent with the radial location and strength of the $m=3$ Fourier components
previously measured at K-band by \citet{rix93}, is strong enough to account for
the angular offset we have measured between the two dominant arms.  
This additional spiral density wave also is consistent with the lopsided  distribution
of molecular gas, and the inter-arm star formation to the north east of the galaxy
nucleus.  Despite molecular gas density variations of a factor of a few  
and corresponding
star formation rate variations (measured from the Pa$\alpha$ emission), 
we see little variation in the computed gas depletion times.
This suggests that the star formation efficiency is not largely affected
by the large gas density variations caused by the presence of an additional
spiral density wave.

Between 1   and 2.2 kpc in M51 we see large scale deviation from
pure $m=2$ or bi-symmetric structure:
an angular offset between the spiral arms, differences in density on one
side compared to the other side, and inter-arm star formation.
All of these three asymmetries are most pronounced in the same radial region 
suggesting that they are related.  We have suggested here that
this phenomena can be explained with one simple unifying model that 
includes an additional weak 3-armed spiral density wave.
The additional wave does not strongly influence the star formation efficiency,
however because of the large gas density variations caused by it,
the wave does influence the star formation rate.  In particular, large
molecular gas densities and associated high star formation rates
are likely to occur where and when the 3-armed and 2-armed
mode both reach their density maxima simultaneously.  We expect
that the 3-armed wave has a different pattern speed (angular rotation rate)
than the 2-armed wave.  Consequently the star formation rate
which is highest when the 3 and 2 armed patterns
constructively add, must be highly variable. 
Even low amplitude waves can cause large changes in 
the gas density.   Despite the low estimated amplitude 
of the 3-armed wave (${\Sigma_3 \over \Sigma_0} \sim 0.05$)
it is likely that it is causing large variations in the molecular
gas density, and consequently in the star formation rate.
The simultaneous propagation of 2 and 3 armed structure
must cause associated large variations in the local star formation rate.
Since our model contains more than one spiral density
wave we would more strongly support scenarios
that include multiple and transient spiral density waves 
(e.g., \citealt{toomre,fuchs}) rather than
those that  focus on a single dominant quasi-steady mode (e.g., \citealt{lowe}).

The model containing two and three armed spiral density waves,
that we have proposed to explain the asymmetries in the central 3kpc of
M51, does not account for all the phenomena we have described in this paper.
For example, this model does not account for the drop in star formation
efficiency observed in the outer parts of spiral arms 1b and 2b between
1.5 and 2 kpc from the nucleus.  It also does not account for
the offsets between the CO and Pa$\alpha$ emission which are largest
at radii greater than 2kpc.
Similar offsets have previously been discussed by \citet{rand99,rand95,rand92,Tilanus91} who
have offered a variety of explanations including the triggering of star formation
by spiral density waves, uneven heating and disassociation by the
resulting star formation and differential clumping of the molecular gas.
However, it will be tempting to search for an additional explanation for these offsets
and the decrease in the star formation
efficiency we see in the outer arms 1b and 2b 
that also involve more than one spiral density wave.
Spiral structure exists in M51 over a large range of radius.  
In section 3.1 we
attempted to fit as large as possible pieces of the spiral structure with as few as 
possible individual components.  However, it is not clear from Figure 1,2 or 3
that the ranges of radius over which we fit individual arm pieces correspond
to separate spiral density waves, traveling at individual pattern speeds.
It is likely that the inner spiral arms of M51 rotate at a different pattern
speed than the outer ones, and the interplay between these waves might provide
an alternative explanation for the offsets observed between the CO and Pa$\alpha$
emission or the low star formation efficiency in the outer parts of spiral arms 1b and 2b.
Future work can further test the possibility that
multiple spiral density waves propagate in galaxies and explore the role
that these multiple modes exert on the distribution of gas and resulting
pattern of star formation.  Future studies
can extend this type study carried out here to more galaxies,
extend the multi-wavelength coverage and exploit the velocity fields that are 
available in the CO and HI data.

\acknowledgments
We thank  
Joel Green, Judith Pipher, Dan Watson, Bruce Elmegreen, Alar Toomre, 
Mousumi Das and Peter Teuben for stimulating discussions.
We are grateful to Tamara Helfer for providing us with access
to the full BIMA SONG galaxy sample data in advance of publication.

\begin{figure*}  
\plotone{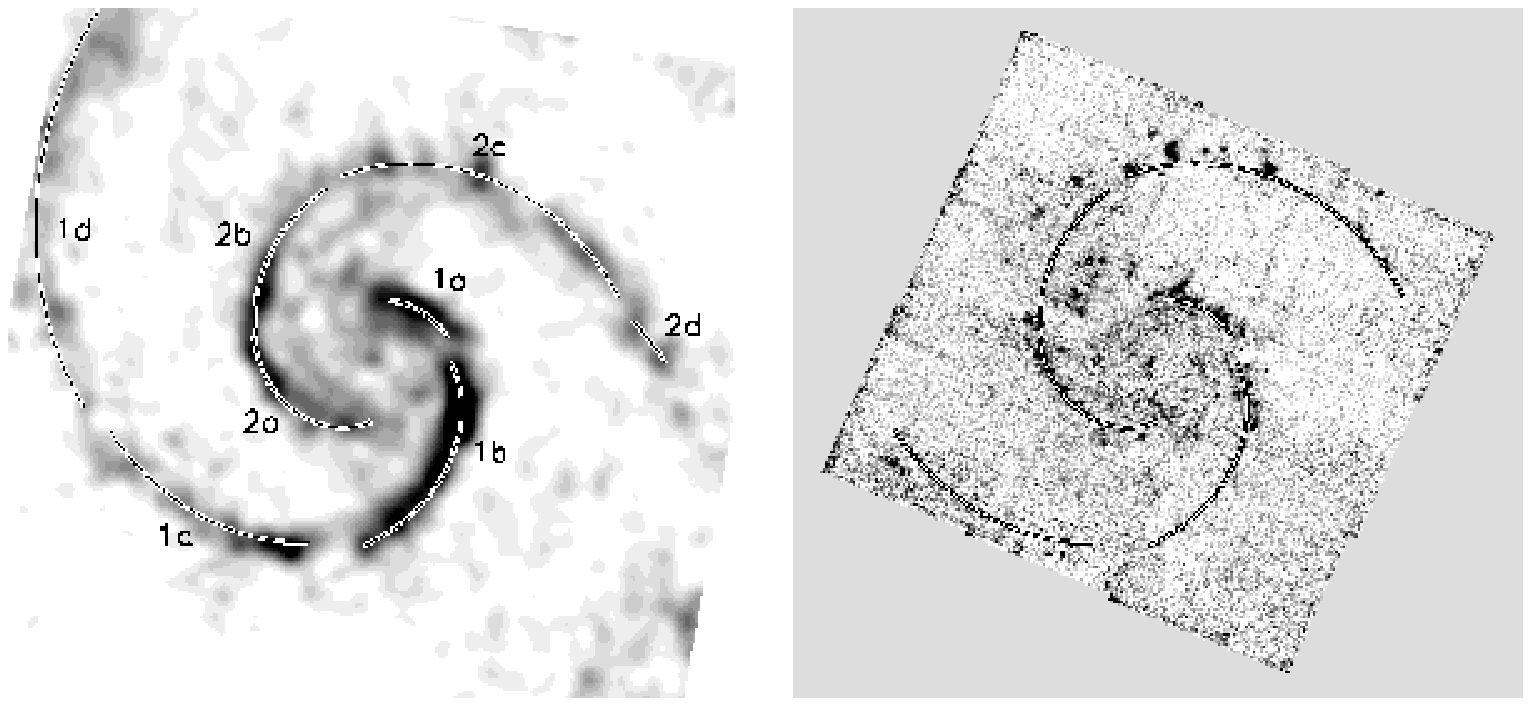} 
\figcaption[ahenry.fig1.eps]{
{\it Left}: We show the BIMA-SONG CO(1-0) image
resampled to be at the same pixel scale and orientation as the 
NICMOS Pa$\alpha$ mosaic ({\it right}). Both images are 
inclination corrected and rotated so that the minor axis is at a P.A. of 170 degrees.  
Overplotted, is our fit to the spiral arms, which was derived from the maxima in the CO image.  Labels correspond to 
different spiral arm regions, each of which was individually 
fit with a logarithmic spiral (equation \ref{logparam}).     
The parameters used to describe them are shown in Table 1. 
Each image is 8.0 kpc wide and 7.6 kpc high.   
Note the inter-arm star formation in the Pa$\alpha$ image NE of the galaxy
nucleus, in between arm 1a and 2b.
}
\end{figure*}

\begin{figure*}  
\epsscale{.7}
\plotone{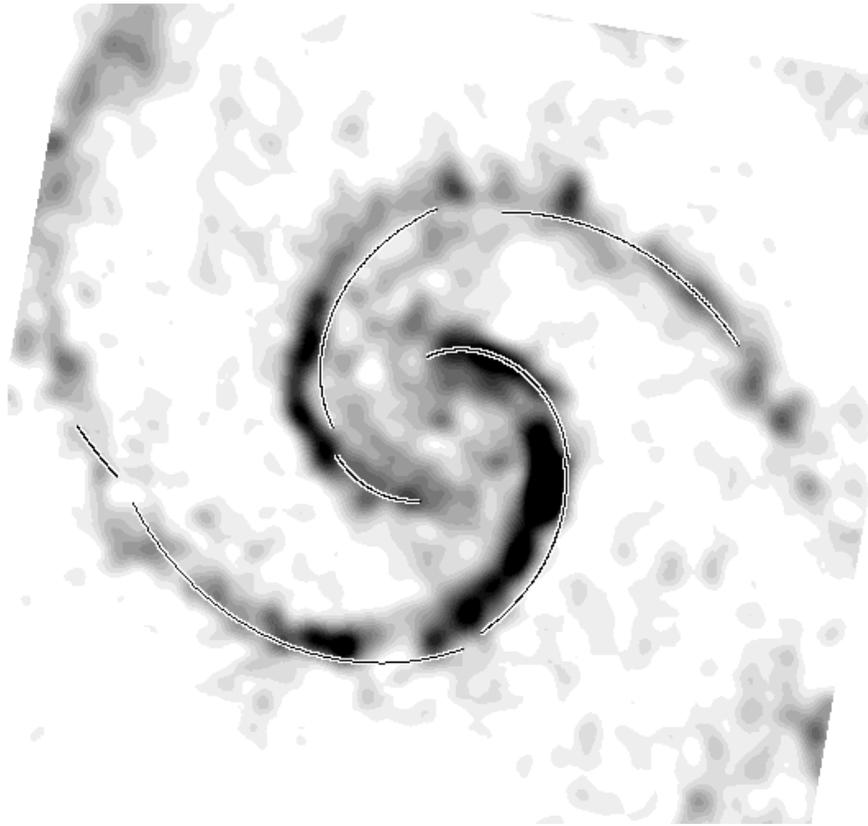}
\figcaption[ahenry.fig2.eps]{In this figure we show the BIMA-SONG CO image, (see Figure 1 left).  
The positions of the over-plotted curves are determined 
by rotating the curves in Figure 1 through an angle of $\pi$.  In the case of $m=2$ symmetry 
the curves would lie on top of the peak CO intensities.  
Asymmetric offsets occur between radii of about 1.2 and 2.4 kpc. }
\end{figure*}

\begin{figure*}
\epsscale{.7}
\plotone{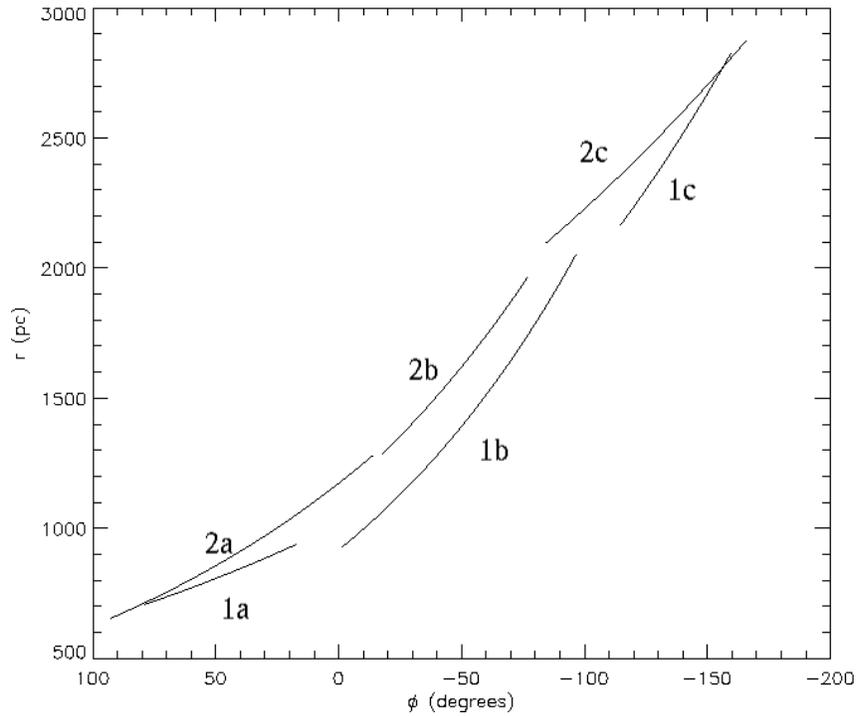}
\figcaption[ahenry.fig3.eps]{In this figure we plot $r$ as a function of $\phi$ 
for spiral arms 1a, 1b, 1c, and for 2a, 2b, and 2c, rotated 
through an angle of $\pi$.  We note a maximum $\Delta \phi$ offset 
of spiral arm 2 from spiral arm 1
of approximately 20 degrees, or 250 pc for the spiral arms labeled 1b and 2b. \label{offsets}}
\end{figure*} 

\begin{figure*} 
\epsscale{1}
\plotone{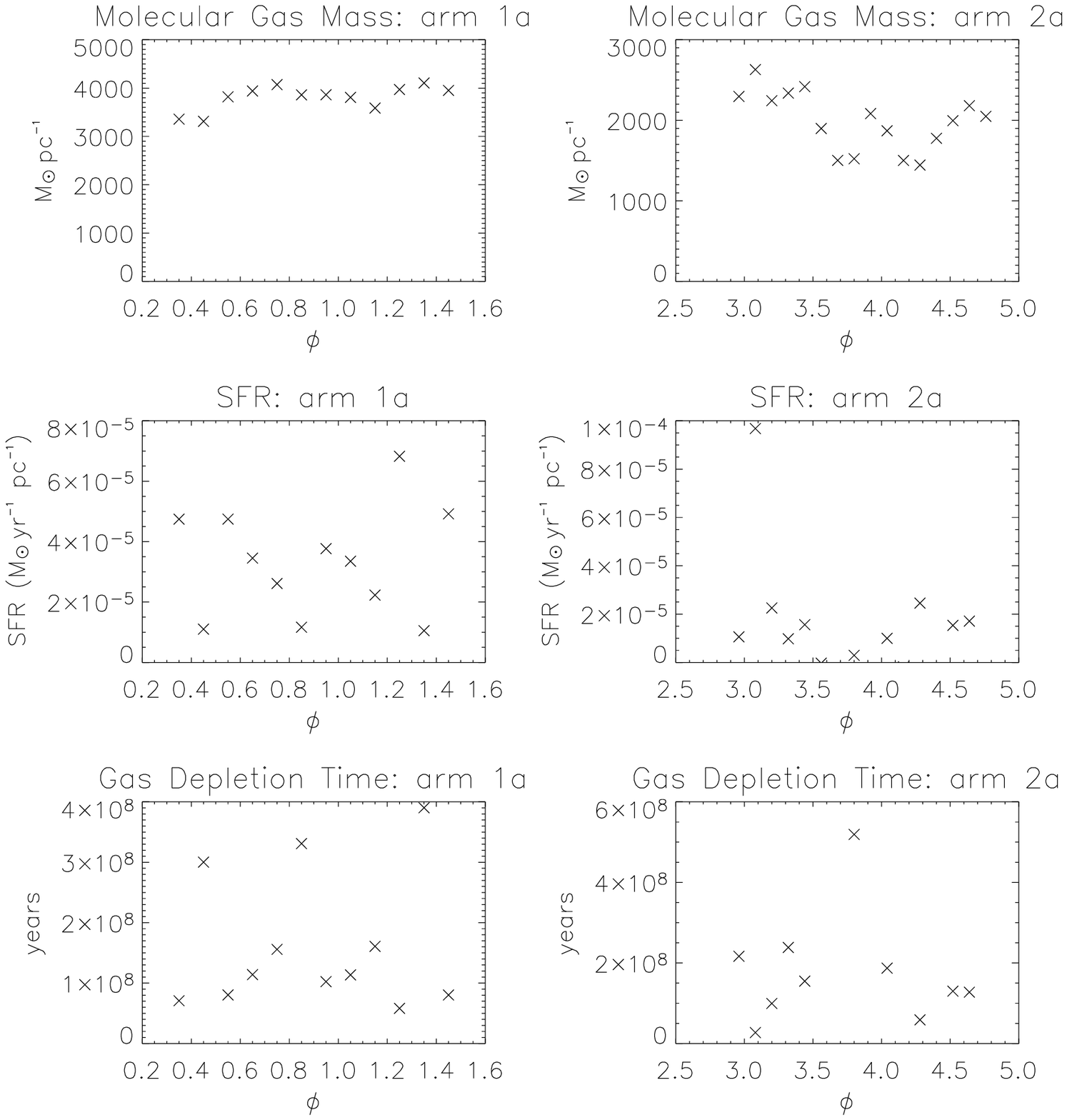} 
\figcaption[ahenry.fig4.eps]{
{\it Top Left}: Molecular gas mass in spiral arm 1a of M51, measured from 
the BIMA-SONG CO image.  
{\it Middle Left}: Star formation rate in spiral arm 1a of M51, as measured from 
the NICMOS Pa$\alpha$ mosaic described in the section 2.  
{\it Bottom Left}: The gas depletion time, in years, derived from the ratio of 
the molecular mass to the star formation rate, in spiral arm 1a.  
{\it Right}: Same as on left, but for spiral arm 2a. 
\label{gdt1}}
\end{figure*}

\begin{figure*}
\plotone{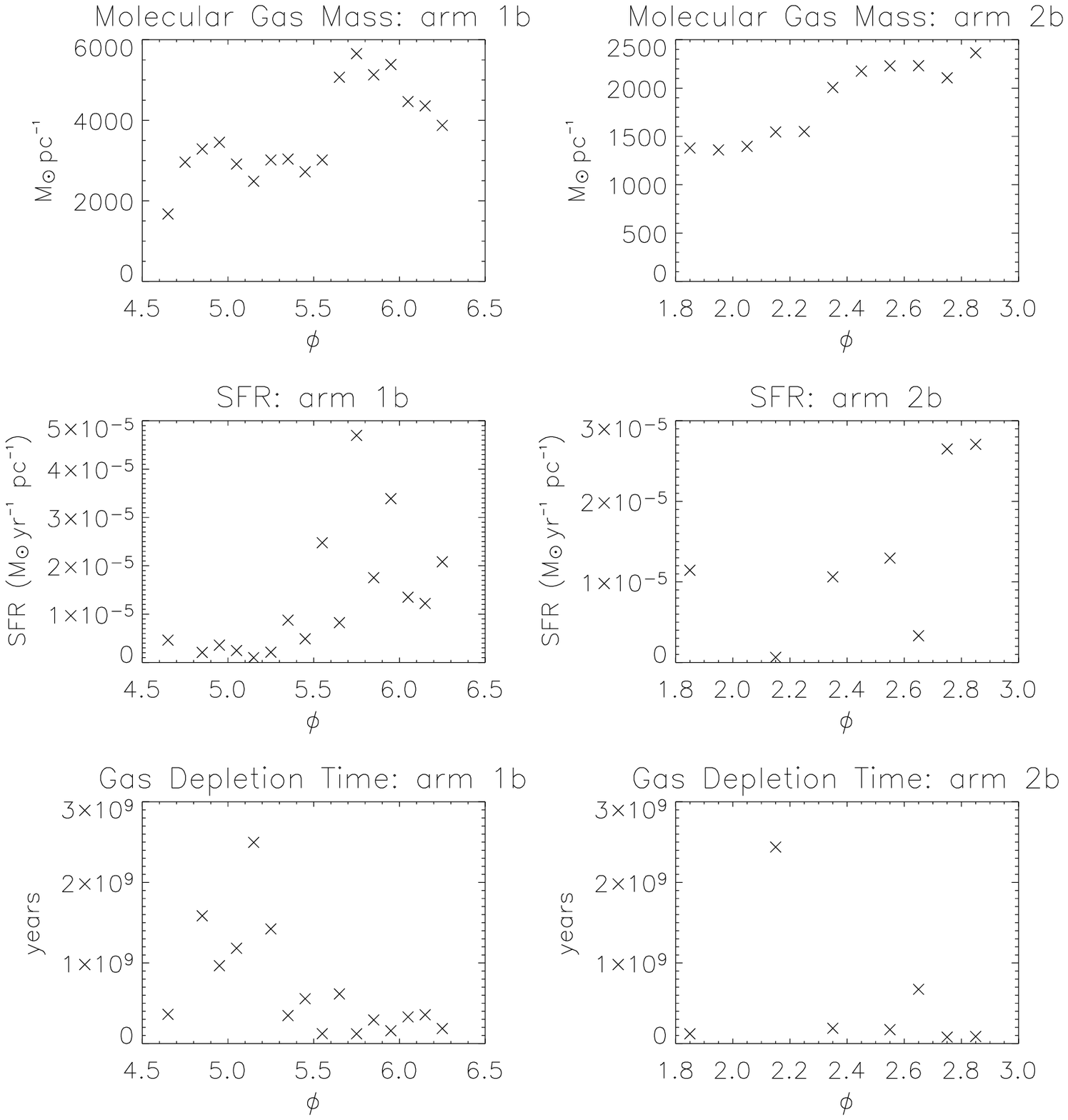}
\figcaption[ahenry.fig5.eps]{
{\it Top Left}: Molecular gas mass in spiral arm 1b of M51, measured from 
the BIMA-SONG CO image.
{\it Middle Left}: Star formation rate in spiral arm 1b of M51, as measured from the NICMOS Pa$\alpha$ 
mosaic described in the section 2.
{\it Bottom Left}: The gas depletion time, in years, derived from the ratio 
of the molecular mass to the star formation rate in spiral arm 1b.
{\it Right}: Same as on left, but for spiral arm 2b.
\label{gdt2}}
\end{figure*}

\begin{figure*}
\epsscale{.5}
\plotone{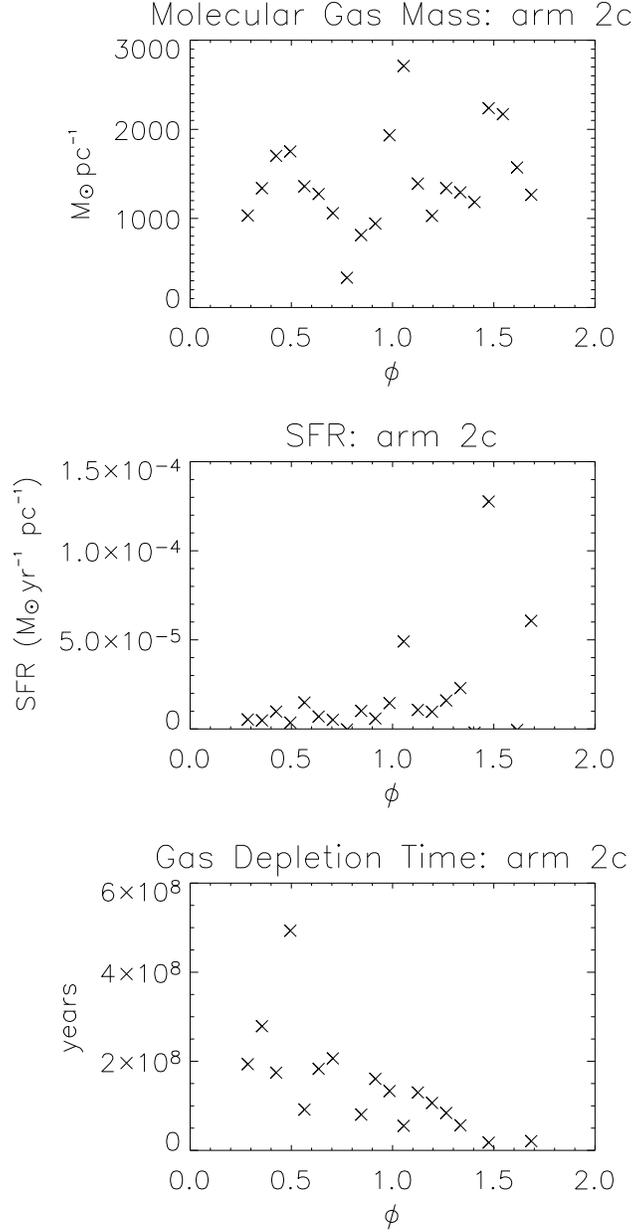}
\figcaption[ahenry.fig6.eps]{
{\it Top}: Molecular gas mass in spiral arm 2c of M51, measured from the BIMA-SONG CO image.
{\it Middle}: Star formation rate in spiral arm 2c of M51, as measured from the NICMOS Pa$\alpha$ mosaic 
described in the section 2.
{\it Bottom}: The gas depletion time, in years, derived from the ratio of 
the molecular mass to the star formation rate in spiral arm 2c. \label{gdt3}}
\end{figure*}

\begin{figure*}
\epsscale{1}
\plotone{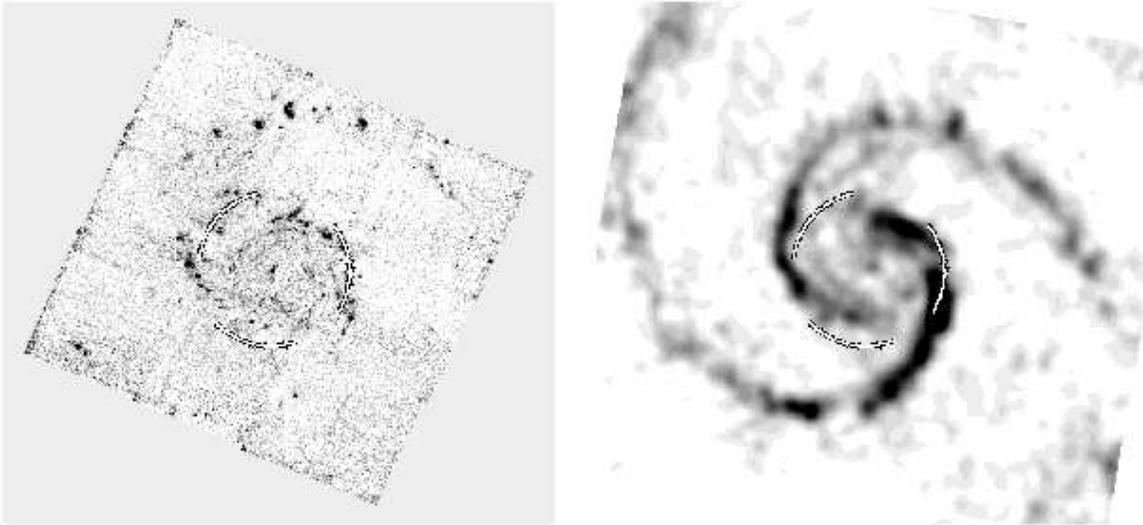}
\figcaption[ahenry.fig7.eps]{
{\it Left}: We show the NICMOS Pa$\alpha$ mosaic, (same as Figure 1, left), with a fit to the inter-arm star formation in the northeast (top left), 
and this same curve, rotated approximately 120 degrees and 240 degrees.
{\it Right}: The same curves, over-plotted on the CO image (same as Figure 1, right). 
\label{tri}}
\end{figure*}

\begin{deluxetable}{cccc}
\tabletypesize{\scriptsize}
\tablecaption{Parameters describing the spiral arms}
\tablewidth{270pt}
\tablehead{
Spiral Arm Region & $\alpha_2$ & $\phi_{0}$ & $\phi$ range  \\ }
\startdata
 1a  & -7.60 & -0.058   &  $0.3 < \phi < 1.4$  \\
 1b  & -4.08 & -6.10    &  $4.6 < \phi < 6.3$  \\
 1c  & -5.88 & -6.55    &  $3.6 < \phi < 4.3$  \\
 1d  & -4.88 & -6.09    &  $2.3 < \phi < 3.3$  \\
 2a  & -5.54 & -3.58    &  $3.0 < \phi < 4.8$  \\
 2b  & -4.86 & -3.44    &  $1.8 < \phi < 2.9$  \\
 2c  & -9.00 & -5.00    &  $0.3 < \phi < 1.7$  \\
 2d  & -2.98 & -1.78    &  $0.0 < \phi < 0.2$  \\
 \enddata
\tablecomments{Logarithmic Spirals are fit to different regions shown in Figure 1.
The parameters $\alpha_2$ and $\phi_{0_2}$ are defined in Equation 2, where
$r$ is in kpc and $\phi$ and $\phi_{0_2}$ are in radians.}

\end{deluxetable}


\begin{thebibliography}{}


\bibitem[Binney \& Tremaine(1987)]{B+T}
Binney, J., \& Tremaine, S.\ 1987,  Galactic Dynamics,
Princeton University Press, Princeton, NJ


\bibitem[Elmegreen et al.(1992)]{elmegreen92}
Elmegreen, B.~G., Elmegreen, D.~M.,  \& Montenegro, L.~1992, ApJS,  79, 37

\bibitem[Fuchs(2001)]{fuchs}
Fuchs, B.~2001, A\&A, 368, 107



\bibitem[Helfer et al.(2003)]{helfer}
Helfer, T.~T., Thornley, M.~D., Regan, M.~W., Wong, T., Sheth, K.,
Vogel, S.~N., Blitz, L., \&  Bock, D.~C.-J.~2003, ApJS, 145, 259

\bibitem[Kennicutt(1989)]{Kennicutt89}
Kennicutt, R.~C. 1989, \apj, 344, 685

\bibitem[Kennicutt(1998)]{Kennicutt98}
Kennicutt, R.~C. 1998, \apj, 498, 541

\bibitem[Kennicutt, Tamblyn, \& Congdon(1994)]{Kennicutt94}
Kennicutt, R.~C., Tamblyn, P., Congdon, C.~E.,  1994, \apj, 435, 22 

\bibitem[Lowe et al.(1994)]{lowe}
Lowe, S.~A., Roberts, W.~W., Yang, J., Bertin, G., Lin, C.~C.~1994,
\apj,  427, 184

\bibitem[Martin \& Kennicutt(2001)]{Martin01}
Martin, C.~L., \& Kennicutt, R.~C. 2001, 555, 301

\bibitem[Osterbrock(1989)]{osterbrock}
Osterbrock, D.E.\ 1989, Astrophysics of Gaseous Nebulae and Active
Galactic Nuclei, University Science Books, Mill Valley, CA

\bibitem[Rand et al.(1992)]{rand92}
Rand, R.~J., Kulkarni, S.~R., \& Rice, W.~1992, \apj, 390, 66

\bibitem[Rand et al.(1999)]{rand99}
Rand, R.~J., Lord, S.~D., \& Higdon, J.L.~1999, \apj, 513, 720

\bibitem[Rand(1995)]{rand95}
Rand, R.~J.~1995, AJ, 109, 2444

\bibitem[Regan et al.(2001)]{regan}
Regan, M.~W., et al.~2001, \apj, 561, 218

\bibitem[Rix \& Rieke(1993)]{rix93}
Rix, H.~W., \& Rieke, M.~J.~1993, \apj, 418, 123

\bibitem[Scoville et al.(2001)]{scoville}
Scoville, N.~Z., Polletta, M., Ewald, S., 
Stolovy, S.~R., Thompson, R. \& Rieke, M.~2001, \apj, 122, 3017

\bibitem[Strong \& Mattox(1996)]{strong}
Strong, A.~W., Mattox, J.~R., 1996, \aap, 308, L21

\bibitem[Tilanus \& Allen(1991)]{Tilanus91}
Tilanus, R.~P.~J., \& Allen, R.~J. 1991, \aap, 244, 8

\bibitem[Toomre(1981)]{toomre}
Toomre, A. 1981~in S.M.~Fall, D.~Lynden-Bell (eds.),
The Structure and Evolution of Normal Galaxies, 
Cambridge Univ. Press, Cambridge, p. 111

\bibitem[Tully(1988)]{Tully}
Tully, R.~B. 1988, Nearby Galaxies Catalog (Cambridge University Press: New York)

\bibitem[Wong \& Blitz(2002)]{Wong02}
Wong, T., \& Blitz, L. 2002, \apj, 569, 157


\end{thebibliography}
\end{document}